# Improved Correlation for Viscosity from Surface Tension Data for Saturated Normal Fluids

**Jianxiang Tian [1, 2, *], Ángel Mulero[3]**

[1]Shandong Provincial Key Laboratory of Laser Polarization and Information Technology Department of Physics, Qufu Normal University, Qufu, 273165, P. R. China

[2]Department of Physics, Dalian University of Technology, Dalian, 116024, P. R. China

[3]Department of Applied Physics, University of Extremadura, Badajoz 06006, Spain

[*]Corresponding author, E-mail address: jxtian@dlut.edu.cn

**Abstract**

*Several correlations between viscosity and surface tension for saturated normal fluids have been proposed in the literature. Usually, they include three or four adjustable coefficients for every fluid and give generally good results. In this paper we propose a new and improved four-coefficient correlation which was obtained by fitting data ranging from the triple point to a point very near to the critical one. Fifty four substances were considered, including simple fluids (such as rare gases), simple hydrocarbons, refrigerants, and some other substances such as carbon dioxide, water or ethanol. The new correlation clearly improves the results obtained with those previously available since it gives absolute average deviations below1% for 40 substances and below 2.1% for 10 substances more.*

## I. Introduction

Surface tension and viscosity are two properties of fluids which are different in nature but whose values need to be known for a wide variety of industrial and physicochemical processes (catalysis, adsorption, distillation, extraction, etc.). The two properties have been extensively studied for normal fluids, and this interest continues.[1-10]

In particular, the viscosity, $\eta$, can be related to the molecular information of the fluid, such as the pair interaction potential function.[2,7] Low-temperature viscosity correlations usually assume that $\ln \eta$ is a linear function of the reciprocal absolute temperature.[1] In the region from about $T_r = T/T_c = 0.7$ (where $T_c$ is the temperature at the critical point) to near the critical point, there are many complex equations available that permit one to express the temperature dependence of viscosity. Examples are the Sastri,[1] Orrick and Erbar,[11,12] and Vogel-Fulcher-Tamman equations.[13,14] This last is the most accurate, and has been widely used.[14,15]

Surface tension, $\sigma$, is also related to the intermolecular interaction potential energy and the liquid interface microstructure.[16,17] It too can be measured with high precision at low and moderate temperatures and pressures. Nevertheless, at high temperatures and high pressures, computer simulations are usually required.[5,6]

Experimental results show that surface tension is a linear function of temperature $T$ for values of $T_r$ between 0.4 and 0.7.[1] At higher temperatures, the surface tension is usually expressed as proportional to one or more terms of the form $(1-T_r)^n$, where $n$ is a fixed constant or substance-dependent coefficient.[1,18-20]

For some fluids, one of these two properties may be more easily measured than the other over certain temperature ranges. Moreover, as indicated previously, both properties are related to the microscopic structure and intermolecular forces of fluids. It is therefore interesting to try to establish some relationship between them. Such a relationship could also be used to check the validity of the measured data, since any deviations may be due to experimental error.[21] Indeed, since both properties are related to the intermolecular potential energy, one might expect there to be some theoretical correlation between the two, although no such link has yet been established.

In 1966, Pelofsky[22] proposed an empirical relationship between the natural logarithm of surface tension and the inverse of viscosity (usually termed the fluidity). Two adjustable coefficients are needed whose values may depend on the temperature range being considered. Queimada et al.[21] checked the use of the Pelofsky correlation for pure compounds and mixtures of n-alkanes, and found adequate results in all cases. The temperature ranges they considered were, however, fairly narrow, and indeed the authors themselves observed that near the critical point the results may be very inaccurate.

More recently, Ghatee et al.[4,8] applied the Pelofsky correlation to some ionic fluids. They found that it was necessary to modify it slightly by introducing an exponent into the viscosity term (we shall denote this hereafter as the modified Pelofsky, or MP, correlation). They initially treated this exponent as an adjustable coefficient, but then they found that its value could be fixed to 0.3 without any significant loss of accuracy for the fluids considered.

Both the original and the modified Pelofsky correlations have recently been studied for a set of 56 normal fluids.[9] It was found that the MP expression gives good results, although, unlike the case for ionic fluids, the corresponding exponent did not take a fixed value.[4,8]

Very recently, Zheng et al.[10] have shown that it is necessary to introduce a fourth adjustable coefficient in order to get a clearly better correlation. This was shown to reproduce the viscosity data available at the NIST Web Book[23] with average absolute deviations below 10% for a set of 40 fluids over a wide range of temperatures.

In the present work, we propose an improved four-coefficient correlation that permits the calculation of the viscosity from knowledge of the surface tension values for normal saturated liquids. In Sec. 2, we describe the new correlation together with the two previous ones.[10] In Sec. 3, we illustrate the results and discuss them. And finally, the conclusions are presented in Sec. 4.

## II. Viscosity/Surface-tension correlations

Pelofsky proposed the following relationship (the P-model) between surface tension and viscosity[22]

$$\ln \sigma = \ln A_1 + \frac{B_1}{\eta} \tag{1}$$

where $A_1$ and $B_1$ are substance-dependent constants. According to that author,[22] this empirical relationship can be applied to both organic and the inorganic pure fluids and mixtures. We have recently studied its accuracy for 56 fluids[9] by calculating the absolute average deviation (AAD) values for the prediction of the surface tension data available in Ref. [23]. We found that the AAD values are less than 2% only for four refrigerants and nonane. Moreover, AADs greater than 20% were found for water, oxygen, and deuterium oxide, for which compounds the P expression is therefore clearly inadequate, at least for the wide temperature range considered.

Those results were improved when the MP expression, proposed by Ghatee et al. for ionic liquids,[4] is used. This expression is:

$$\ln \sigma = \ln C + D \left( \frac{1}{\eta} \right)^{\phi} \tag{2}$$

where $C$, $D$, and the exponent $\phi$ are substance-dependent coefficients. We have found[9] that this correlation improves the results significantly for 34 out of the 56 fluids considered.

As in the present paper we are interested in the calculation of the viscosity, we shall use the alternative form of Eq. (2):

$$\left(\frac{1}{\eta}\right)^\phi = A_2 + B_2 \ln \sigma \tag{3}$$

Here $A_2$ and $B_2$ are substance-dependent coefficients. We have recently shown[10] that this form of the MP expression can reproduce the viscosity data in Ref. [23] for 36 fluids (out of 40 considered) with AADs below 10%.

To improve this result, we proposed a new four-parameter correlation, which we called the **ZTM4** correlation, given by[10]

$$\ln \eta = A_3 + \frac{B_3}{\sigma^{\frac{1}{n}} + C_3} \tag{4}$$

Here $n$, $A_3$, and $B_3$ are substance-dependent coefficients. By using the ZTM4 correlation, it is possible to obtain AADs below 10% for the set of 40 fluids considered, being lower than 1% for 5 of them. Only for six fluids were AAD values greater than 5% found.

Unfortunately, in the particular cases of R13, isobutene, and propane, none of the above models can give AAD values lower than 5%. There is therefore some room for improvement by developing new correlation models connecting the viscosity and the surface tension of fluids, and this is the purpose of the present paper. At the same time, it is known that the data presently available in Ref. [23], which we used as references, have been not updated for a long time. So in this paper we have used instead those available in the last version of the REFPROP program by NIST.[24] This permits to update the results given in Ref. [10].

To derive a new correlation that permits the viscosity to be obtained from the surface tension, we reconsidered the main idea of Pelofsky which was to consider the natural logarithm of the surface tension instead of the viscosity. Thus, we introduce a new correlation as follows:

$$\ln \sigma = A + \frac{B}{\eta^m + C} \tag{5a}$$

with the alternative form being

$$\frac{1}{\eta^m + C_4} = A_4 + B_4 \ln \sigma \tag{5b}$$

or

$$\eta = \left( \frac{1}{A_4 + B_4 \ln \sigma} - C_4 \right)^{1/m} \qquad (5c)$$

where $A_4$, $B_4$, $C_4$, and $m$ are adjustable coefficients which have to be determined by using an adequate set of data for both properties, the surface tension and the viscosity, available over the same range of temperatures. In this paper, the units of surface tension and viscosity are given in *N/m* and *cP*, respectively.

## III. Results and Discussion

As the main objective of the present work was to study the relationship between two properties, it was important to adequately select the source of the data used to this end. We thus selected the REFPROP program[24] because the data it offers are sufficiently accurate and are straightforwardly available. Fifty four fluids were picked. These included simple fluids (such as argon and other rare gases), simple hydrocarbons, refrigerants, and some other substances (carbon dioxide, water, ethanol, etc.). These substances are listed in Table I, in alphabetical order for three kinds of substances: refrigerants, hydrocarbons, and other common fluids. The data start at the temperature $T_0$, which is the triple point temperature except for R14, parahydrogen, ethanol and $CO_2$ (for which no data are available below $T_0$ in REFPROP), and finish at the temperature $T_f$, which is near the critical point. The small differences between $T_f$ and the critical temperature, $T_c$, can be observed by comparing the last two columns of Table I. To have an adequate number of data, most of them were obtained with a temperature increment of 1 *K*, but in those cases for which the temperature range is short the increment was 0.5 *K*.

The data for the surface tension and the viscosity were used to check the behaviour of the previous MP and ZTM4 correlations as well as the new one proposed here, Eq. (5c). During the fitting procedure, those coefficients that minimized the AAD values were chosen. The coefficients for the new, the MP, and the ZTM4 expressions are available as supplementary material.

To calculate the AAD, we first calculated the percentage deviation (PD) between the values for the viscosity obtained from the correlation by introducing the surface tension as input $\eta(\sigma_i)$, and the data offered by REFPROP [24], $\eta_i$, as follows:

$$\text{PD}_i = 100\left(\eta(\sigma_i) - \eta_i\right)/\eta_i, \qquad (6)$$

A positive $\text{PD}_i$ value means that the model overestimates the accepted datum, whereas a negative $\text{PD}_i$ value means that the model underestimates it. Then we calculated the average absolute percentage deviation for every fluid:

$$\text{AAD} = \frac{\sum_{i=1}^{N}|\text{PD}_i|}{N} \quad (\%), \qquad (7)$$

where $N$ is the number of data.

It has to be borne in mind that, since AAD is a percentage, it is influenced by the high individual PD values that can be found when the viscosity takes very low values (near to zero), which occurs at the highest temperatures, i.e., near the critical point temperature. This means that, near the critical point, the absolute deviations are low, but the relative PD can take very high values, and this has a clear influence on the final AAD value.

The AAD values obtained for the three correlations analyzed are given in Table I. In Table II the results are summarized by giving the number of fluids for which each correlation gives an AAD value lower or higher than a given quantity. As can be seen, the newly proposed expression gives AADs below 1% for 40 out of the 54 fluids, whereas for the rest of the fluids the AAD ranges from 1% to 4.6%. In general, while the ZTM4 correlation can give AADs below 1% for just 8 fluids, it gives AADs greater than 5% for 9 fluids – two refrigerants and seven hydrocarbons (see Table I). Finally, as the MP model has one less adjustable coefficient, it consequently gives poorer results. In particular, AADs greater than 10% are found for nine fluids.

In comparison with the overall results obtained from the MP and ZTM4 correlations, the use of the new one leads to a clear overall improvement, with only a

few exceptions:

(i) In the case of methane and xenon the new correlation gives very low values (0.23% and 0.29%, respectively), but the ZTM4 gives similar although lower values.

(ii) For methanol both the ZTM4 and the new correlation give high and very similar AAD values;

(iii) In the case of dimethylether the ZTM4 gives a better result in general. As can be seen in Fig. 1, the main drawback with the new correlation is that it gives very high PD values near the triple point (high surface tension values). On the other hand, despite the ZTM4 gives a lower AAD value, it gives high PD values near the critical point (surface tension values near to zero).

Let us consider various examples illustrating the behaviour of the relationship between the viscosity and the surface tension and the accuracy of the correlations studied.

As can be seen in Table I, in the case of refrigerant substances our new expression gives AADs below 1.1%. For instance, a clear improvement is found in the case of R13 when Eq. (5c) is used instead of the previous correlations. Fig. 2 shows that, for low surface tension values, the absolute deviations are small, whereas Fig. 3 shows that the relative PDs are high in the case of the MP and ZTM4 correlations. For high surface tension values, the absolute and relative deviations are high for both the MP and ZTM4 correlations, and are clearly lower in the case of the new correlation proposed here.

In general, the success of the new correlation does not depend on the range of values of the surface tension or viscosity. This can be seen in Fig. 4 for R143a, where the values of viscosity and surface tension are different to that shown in Fig. 2, and where it is shown that the new correlation gives very adequate results over the whole temperature range considered.

In the case of hydrocarbons, the new correlation clearly improves the results obtained with the previous expressions, with the above mentioned exception of methane, and also in the case of isopentane, for which none of the three expressions can give an AAD below 4.5%. As can be seen in Fig. 5, this is due to the sharp increasing of the surface tension values when the temperature decreases (at low temperatures). Although it is not appreciated in the figure, the PD given by the ZTM4 correlation at high temperatures (low surface tension values) are high. The ZTM4 correlation behaves similarly for other hydrocarbons, and it cannot be considered as an adequate correlation for this kind of fluids, because it gives AADs below 2% only for 3 out of the 18 hydrocarbons considered.

We shall now analyse the results for the other fluids considered. The AADs obtained with the new correlation, Eq. (5c), are below 0.7% except for dimethylether (see Fig. 1 and comments above), methanol, hydrogen sulfide, and water.

In the case of methanol, the behavior of the viscosity-surface tension curve is very similar to those shown in Fig. 5, so the obtained results are very similar to those for isopentane, and none of the correlations can give an AAD below 3.6%.

In the case of hydrogen sulfide, the best result for the new correlation is found when $C_4 = 0$, and therefore this corresponds to the MP expression. As can be seen in Fig. 6, the worst results are obtained at the lowest and highest temperatures (near the triple and critical points).

Finally, we shall consider the case of water, for which none of the correlations used can give AADs below 3.3%. For this fluid the viscosity-surface tension curve Fig. 7 is similar to that of isopentane, i.e. the viscosity values are very small for a large range of high temperatures (low surface tension values), but sharply increased at low temperatures (high surface tension values). Indeed, as can be seen in Fig. 7, none of the correlations can adequately reproduce the trend of the data at either low or high temperatures.

## IV. Conclusions

Three models for the correlation between viscosity and surface tension have been checked for fifty four fluids of different kinds. Data from REFPROP[24] were considered as references. The results for the viscosity data were tested by obtaining percentage deviations for every datum and the absolute average deviation for each fluid.

The first correlation was the MP expression,[4] which has three adjustable coefficients. It gives AADs below 5% only for 25 out of the 54 fluids considered. For R14, dodecane, ethene, nonane, pentane, propane, carbon monoxide, hydrogen, hydrogen sulfide, and krypton, it gives better results than the ZTM4 correlation. In the case of hydrogen sulfide it gives the same result as the new correlation considered here, just because the best fit to this last is obtained with a less coefficient. On the other hand, MP gives AADs greater than 10% for 9 fluids.

The second correlation is the recently proposed ZTM4 expression.[10] This gives AADs below 5% for 45 fluids and below 10% for all of them except propane. For methane, dimethylether, methanol, xenon this correlation gives better results than the new expression proposed here, although in the case of methane and methanol the difference is practically inappreciable.

Finally, the third correlation is that proposed here, Eq. (5c). It also contains four adjustable coefficients. We have shown that it yields the lowest overall deviations for 50 of the substances considered, and absolute average deviations below 5% for all of them. In particular, it gives AAD values below 1.1% for the 20 refrigerants, below 2.1% for 17 out of the 18 hydrocarbons, and below 1.5% for 13 out of the 16 other substances. Details are shown in Tables I and II.

Although there might be some room for improvement by developing new correlation models connecting the viscosity and the surface tension of fluids, the new correlation proposed here is easy to use and highly accurate (at least for the fluids studied here). It should therefore be considered in the future to study other kinds of fluids for which only more limited data are available.

**Acknowledgements**

The National Natural Science Foundation of China under Grant No. 11274200, the Natural Science Foundation of Shandong Province under Grant No. ZR2011AM017, and the foundations of QFNU and DUT have supported this work (J.T.). It was also partially supported by the "Gobierno de Extremadura" and the European Union (FEDER) through project GR10045 (A.M.).

**Supporting Information**

The coefficients for the new, the MP, and the ZTM4 expressions are available as supplementary material. This information is available free of charge via the Internet at http://pubs.acs.org/.

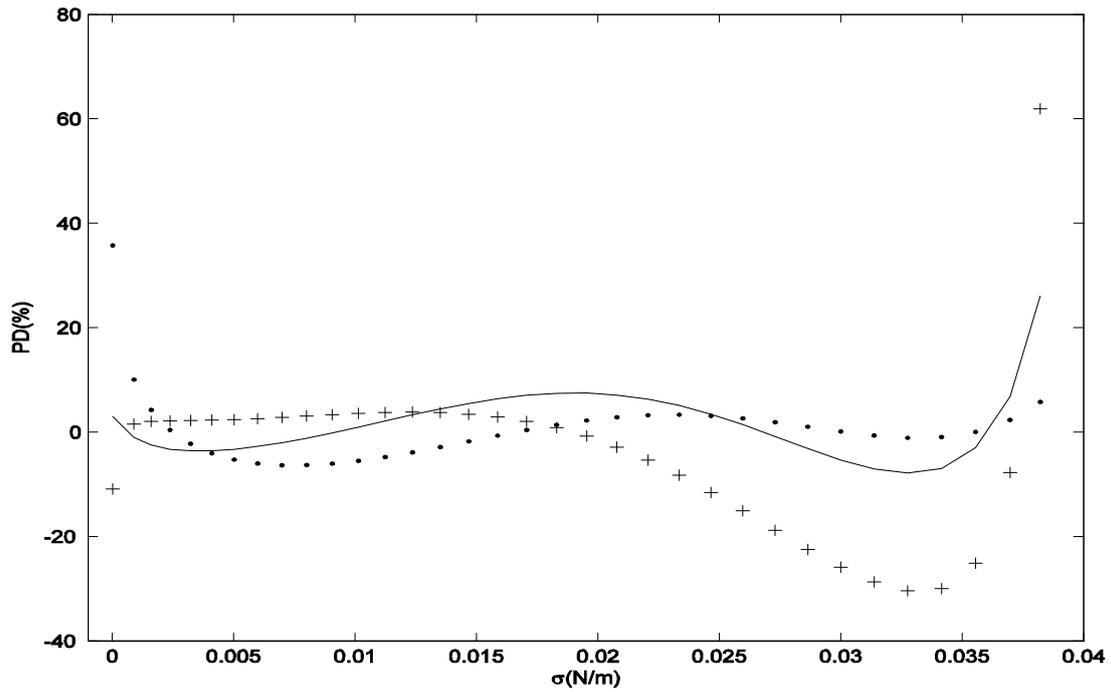

**Figure 1.** Percentage deviations for the calculation of the viscosity for dimethylether from three equations. dots: ZTM4; crosses: MP; continuous line: Eq. (5c).

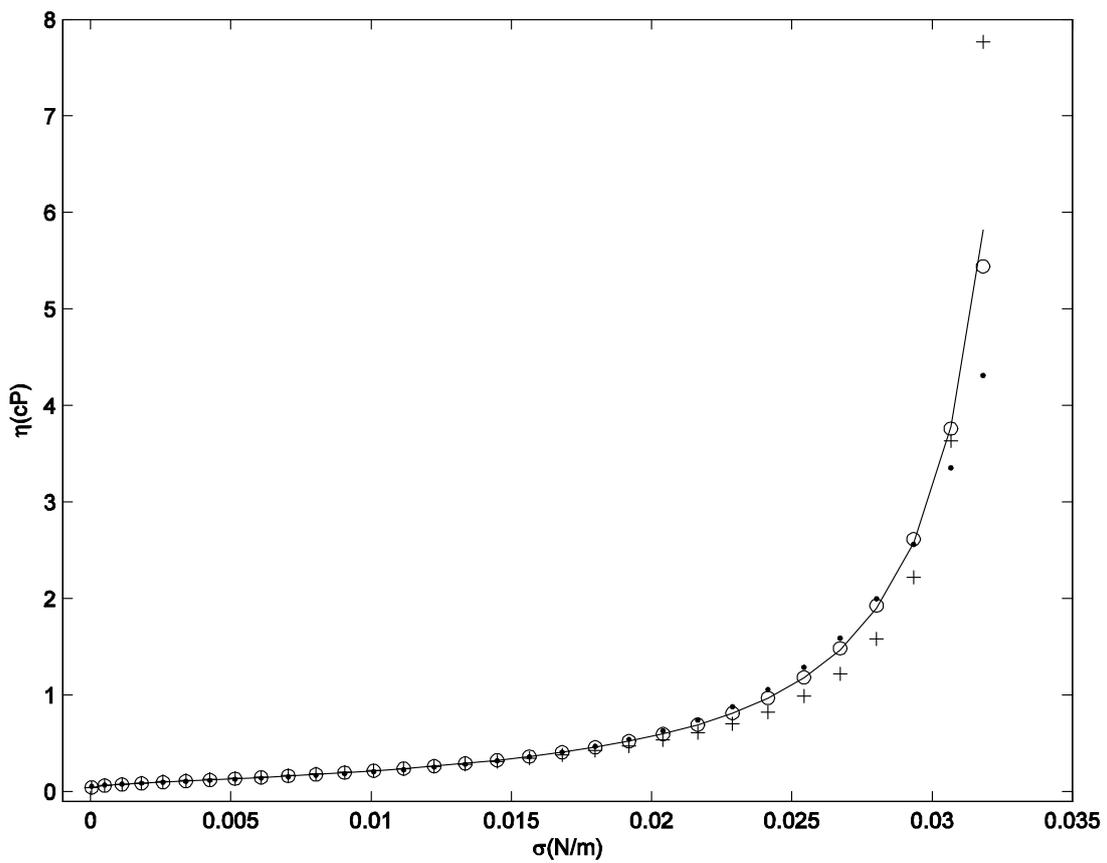

**Figure 2.** Viscosity versus surface tension for R13. Circles: NIST data; dots: ZTM4; crosses: MP; continuous line: Eq. (5c).

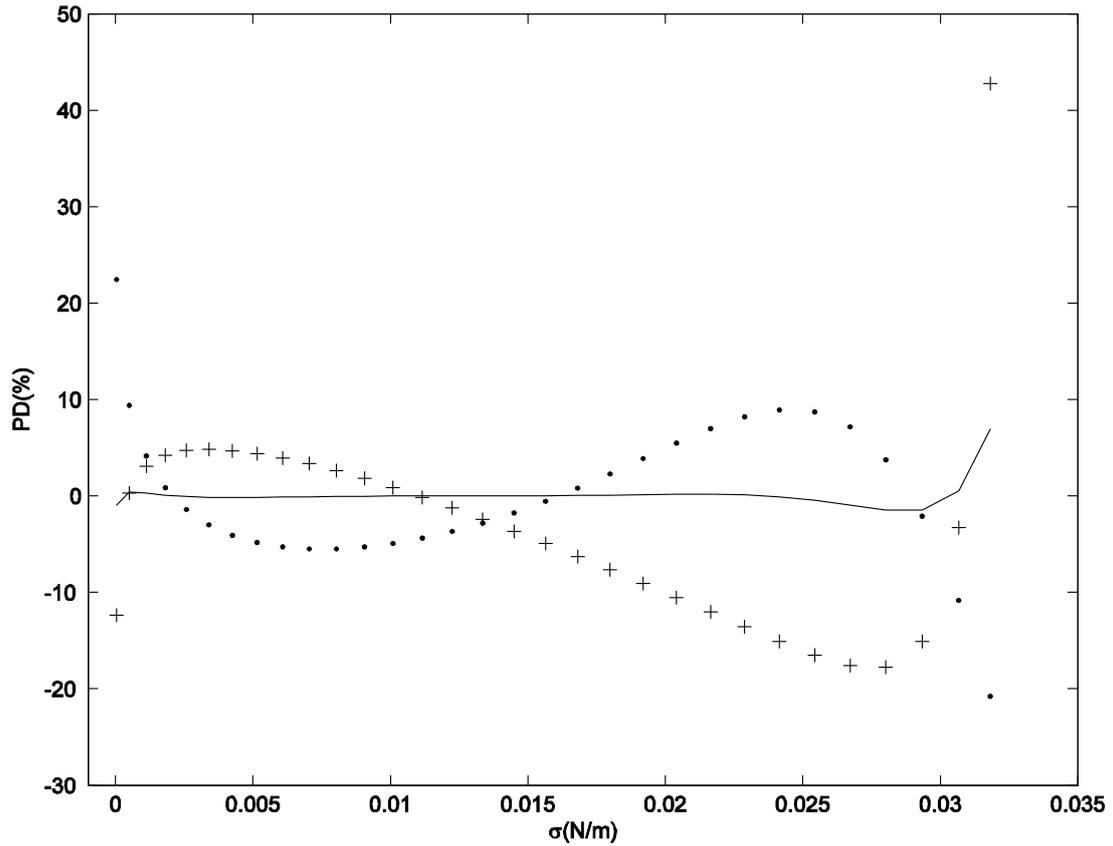

**Figure 3.** Percentage deviations for the calculation of the viscosity for R13 from three equations. dots: ZTM4; crosses: MP; continuous line: Eq. (5c).

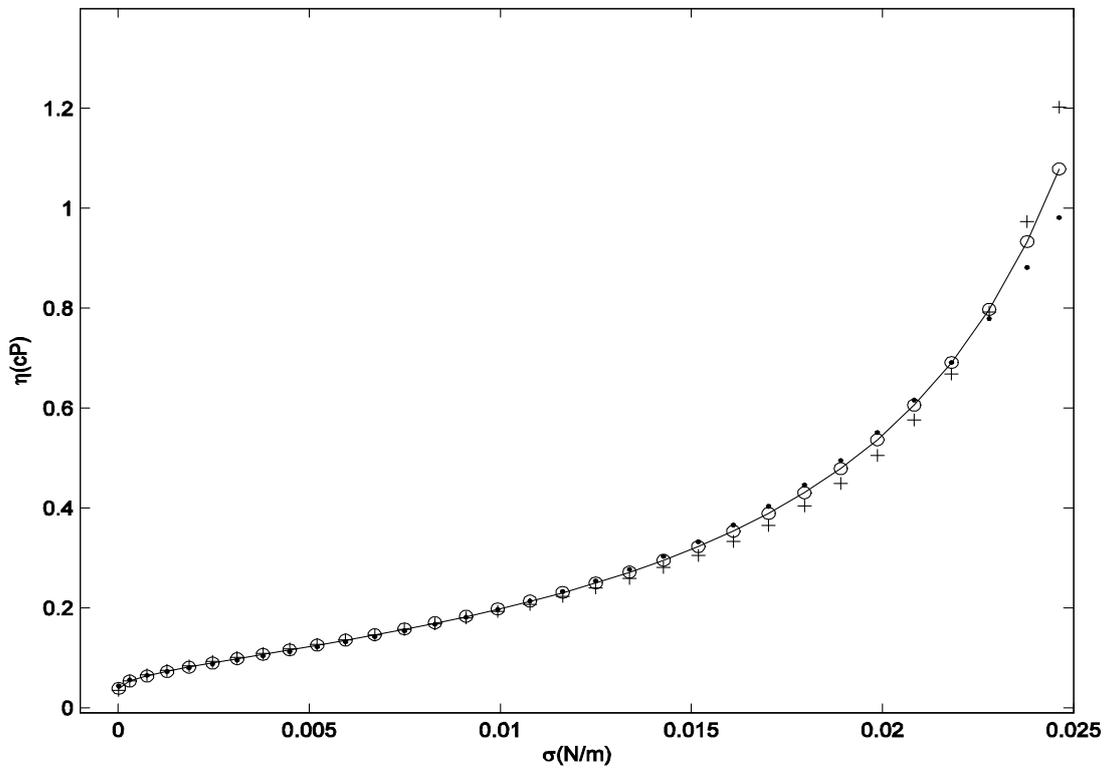

**Figure 4.** Viscosity versus surface tension for R143a. Circles: NIST data; dots: ZTM4; crosses: MP; continuous line: Eq. (5c).

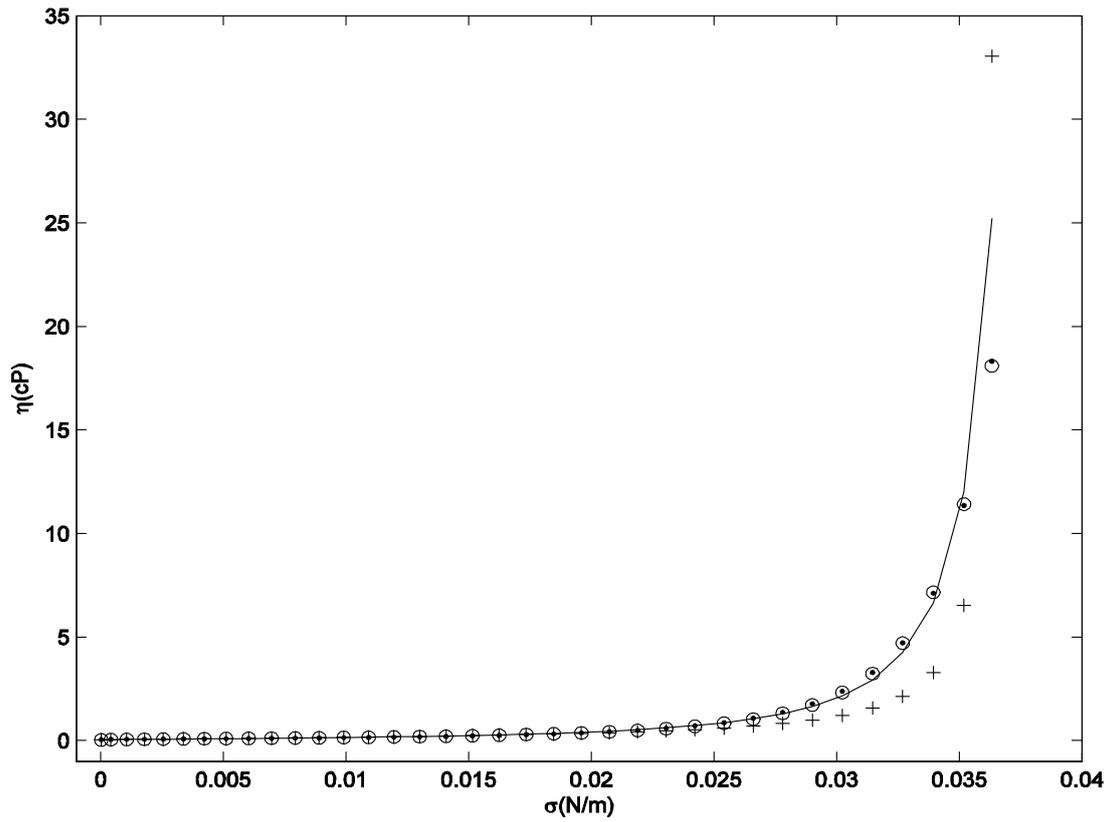

**Figure 5.** Viscosity versus surface tension for isopentane. Circles: NIST data; dots: ZTM4; crosses: MP; continuous line: Eq. (5c).

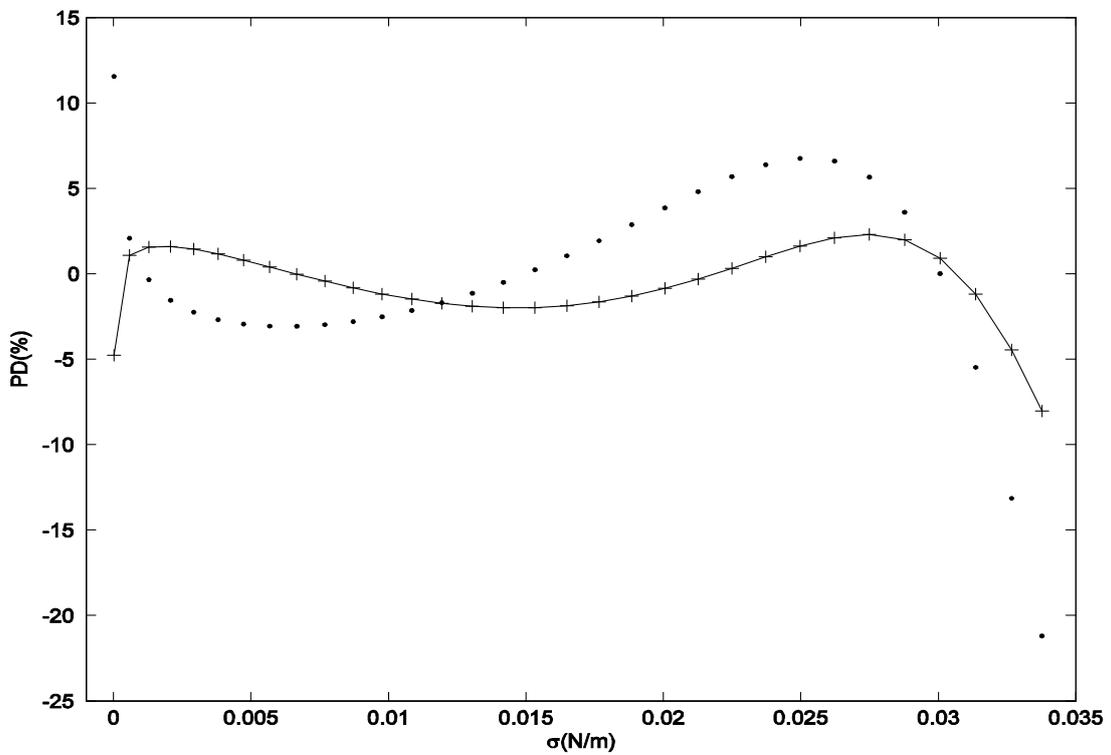

**Figure 6.** Percentage deviations for the calculation of the viscosity for hydrogen sulfide from three equations. Dots: ZTM4; crosses: MP; continuous line: Eq. (5c).

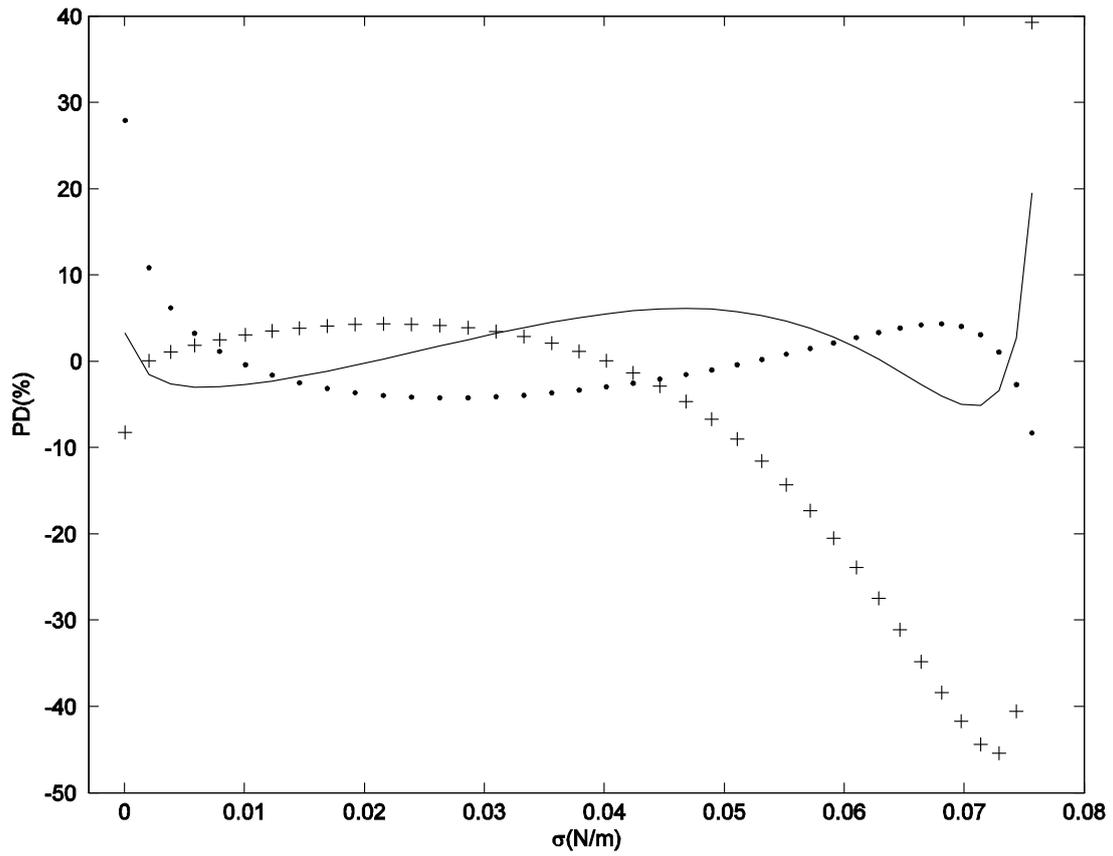

**Figure 7.** Percentage deviations for the calculation of the viscosity for water from three equations. Dots: ZTM4; crosses: MP; continuous line: Eq. (5c)

Table I. AADs (%) of the viscosity for the following correlations: new, Eq. (5c); ZTM4, Eq. (4); MP, Eq. (3). For all the substances except R14, carbon dioxide, parahydrogen and ethanol (for which no data are available in the range of triple point in Ref. [24]), the initial temperature $T_0$ is equal to the triple point temperature $T_{tr}$. The final temperature $T_f$ is the temperature near the critical point $T_c$. The lowest AAD values for every fluid are in boldface. For substances marked by star, the temperature increment is $0.5K$. For others, the temperature increment is $1K$.

| substances | AAD NEW | AAD ZTM4 | AAD MP | $T_{tr}$ (K) | $T_o$ (K) | $T_c$ (K) | $T_f$ (K) |
|---|---|---|---|---|---|---|---|
| REFRIGERANTS | | | | | | | |
| R13 | **0.35** | 5.18 | 7.21 | 92.00 | 92.00 | 302 | 301 |
| R14* | **0.17** | 1.65 | 0.75 | 89.54 | 120 | 227.51 | 227 |
| R22 | **0.94** | 7.96 | 8.33 | 115.73 | 115.73 | 369.3 | 368.73 |
| R23 | **0.94** | 4.00 | 6.90 | 118.02 | 118.02 | 299.29 | 299.02 |
| R32 | **0.19** | 1.69 | 4.86 | 136.34 | 136.34 | 351.26 | 350.34 |
| R41 | **0.30** | 4.17 | 4.62 | 129.82 | 129.82 | 317.28 | 316.82 |
| R123 | **0.74** | 2.76 | 8.82 | 166.00 | 166.00 | 456.83 | 456.00 |
| R125 | **0.34** | 1.71 | 3.07 | 172.52 | 172.52 | 339.17 | 338.52 |
| R134a | **0.44** | 3.75 | 4.37 | 169.85 | 169.85 | 374.21 | 373.85 |
| R141b | **0.58** | 4.47 | 6.48 | 169.68 | 169.68 | 477.5 | 476.68 |
| R142b | **0.24** | 4.01 | 7.84 | 142.72 | 142.72 | 410.26 | 409.72 |
| R143a | **0.31** | 2.51 | 3.42 | 161.34 | 161.34 | 345.86 | 345.34 |
| R152a | **0.14** | 2.69 | 6.10 | 154.56 | 154.56 | 386.41 | 385.56 |
| R1234yf | **0.19** | 0.95 | 1.76 | 220.0 | 220.0 | 367.85 | 367 |
| R1234ze(E) | **0.34** | 3.45 | 3.61 | 168.62 | 171.62 | 382.51 | 381.62 |
| R218 | **0.79** | 4.80 | 10.4 | 125.45 | 125.45 | 345.02 | 344.45 |
| R227ea | **0.65** | 3.02 | 6.24 | 146.35 | 146.35 | 374.9 | 374.35 |
| R245fa | **1.07** | 4.02 | 9.69 | 171.05 | 171.05 | 427.16 | 427.05 |
| R717 | **0.26** | 2.54 | 3.04 | 195.5 | 195.5 | 405.4 | 404.5 |
| RC318 | **0.72** | 0.96 | 5.77 | 233.35 | 233.35 | 388.38 | 388.35 |
| HYDROCARBONS | | | | | | | |
| Butane | **0.29** | 4.18 | 7.06 | 134.90 | 134.9 | 425.13 | 424.9 |
| Cyclohexane | **0.23** | 1.85 | 5.75 | 279.47 | 279.47 | 553.6 | 553.47 |
| Decane | **0.65** | 4.38 | 5.76 | 243.50 | 243.5 | 617.7 | 617.5 |
| Dodecane | **1.04** | 5.18 | 4.09 | 263.60 | 263.6 | 658.1 | 657.6 |
| Ethane | **0.47** | 4.33 | 4.96 | 90.368 | 90.368 | 305.32 | 304.37 |
| Ethene | **0.62** | 3.60 | 2.60 | 103.99 | 103.99 | 282.35 | 281.99 |
| Heptane | **1.79** | 4.99 | 8.07 | 182.55 | 182.55 | 540.13 | 539.55 |
| Hexane | **1.50** | 3.51 | 7.28 | 177.83 | 177.83 | 507.82 | 506.83 |
| Isobutane | **1.59** | 5.77 | 16.7 | 113.73 | 113.73 | 407.81 | 407.73 |
| Isopentane | **4.56** | 4.89 | 16.5 | 112.65 | 112.65 | 460.35 | 459.65 |
| Methane | 0.23 | **0.22** | 3.66 | 90.694 | 91.694 | 190.56 | 189.69 |
| Nonane | **0.64** | 7.55 | 2.01 | 219.70 | 219.7 | 594.55 | 593.7 |
| Octane | **0.52** | 3.90 | 6.96 | 216.37 | 216.37 | 569.32 | 568.37 |
| Pentane | **1.50** | 7.84 | 5.46 | 143.47 | 143.47 | 469.7 | 469.47 |
| Propane | **1.66** | 11.7 | 10.3 | 85.525 | 85.525 | 369.89 | 369.53 |
| Propanone | **0.40** | 1.67 | 14.7 | 178.50 | 178.50 | 508.10 | 507.5 |
| Propene | **2.02** | 7.10 | 14.5 | 87.953 | 87.953 | 364.21 | 363.95 |
| Toluene | **1.83** | 6.63 | 13.9 | 178.0 | 178.00 | 591.75 | 591 |
| OTHERS | | | | | | | |
| Argon | **0.28** | 0.86 | 1.38 | 83.806 | 83.806 | 150.69 | 149.81 |
| Carbon dioxide | **0.03** | 0.05 | 1.90 | 216.59 | 217.09 | 304.13 | 303.59 |
| Carbon monoxide | **0.49** | 2.29 | 0.65 | 68.16 | 68.16 | 132.86 | 132.16 |
| Dimethylether | 4.28 | **3.80** | 9.03 | 131.66 | 131.66 | 400.38 | 399.66 |
| Ethanol | **0.56** | 0.78 | 9.64 | 159.0 | 250 | 514.71 | 514 |
| Hydrogen* | **0.65** | 1.12 | 0.80 | 13.957 | 13.957 | 33.145 | 32.457 |
| Hydrogen sulfide | **1.47** | 3.70 | **1.47** | 187.70 | 187.7 | 373.1 | 372.7 |
| Krypton | **0.14** | 1.34 | 1.01 | 115.78 | 115.78 | 209.48 | 208.78 |
| Methanol | 3.70 | **3.68** | 23.4 | 175.61 | 175.61 | 512.60 | 511.61 |
| Nitrogen | **0.53** | 1.81 | 4.93 | 63.151 | 63.151 | 126.19 | 126.15 |
| Nitrogen Oxide | **0.24** | 1.19 | 2.48 | 182.33 | 182.33 | 309.52 | 309.33 |
| Oxygen | **0.58** | 2.53 | 5.62 | 54.361 | 54.361 | 154.58 | 154.36 |
| Parahydrogen* | **0.51** | 1.10 | 1.52 | 13.803 | 13.903 | 32.938 | 32.803 |
| Sulfur Hexafluoride | **0.33** | 0.33 | 1.78 | 223.56 | 223.56 | 318.72 | 317.56 |
| Water | **3.39** | 3.49 | 12.77 | 273.16 | 273.16 | 647.1 | 646.16 |
| Xenon | 0.29 | **0.22** | 3.22 | 161.41 | 161.41 | 289.73 | 289.41 |

**Table II.** Number of fluids satisfying different AAD ranges to verify the capacities of correlations

| AAD range | MP | ZTM4 | Equation (5c) |
|---|---|---|---|
| <0.5% | 0 | 4 | 24 |
| <1% | 3 | 8 | 40 |
| <2% | 10 | 18 | 49 |
| <5% | 25 | 45 | 54 |
| <10% | 45 | 53 | 54 |
| >10% | 9 | 1 | 0 |